\documentclass[twocolumn,showpacs,preprintnumbers,amsmath,amssymb]{revtex4}
\usepackage{graphicx}
\usepackage{dcolumn}
\usepackage{bm}

\begin{document}

\preprint{APS/123-QED}

\title{Comment on ``Quantum secret sharing based on reusable Greenberger-Horne-Zeilinger states as secure carriers'' [Phys. Rev. A 67, 044302 (2003)]}

\author{Fei Gao$^{1,2}$, \quad Fenzhuo Guo$^{1}$, \quad Qiaoyan Wen$^{1}$, and Fuchen Zhu$^{3}$\\
        (1. School of Science, Beijing University of Posts and Telecommunications, Beijing, 100876, China) \\
        (2. State Key Laboratory of Integrated Services Network, Xidian University, Xi'an, 710071, China)\\
        (3. National Laboratory for Modern Communications, P.O.Box 810, Chengdu, 610041, China)\\ Email: hzpe@sohu.com}

\date{\today}

\begin{abstract}
In a recent paper [S. Bagherinezhad and V. Karimipour, Phys. Rev.
A \textbf{67}, 044302 (2003)], a quantum secret sharing protocol
based on reusable GHZ states was proposed. However, in this
Comment, it is shown that this protocol is insecure if Eve employs
a special strategy to attack.
\end{abstract}


\maketitle

In a recent paper \cite {BK}, Bagherinezhad and Karimipour
proposed a quantum secret sharing protocol based on reusable GHZ
states. The security against both intercept-resend strategy and
entangle-ancilla strategy was proved. However, here we will show
that with the help of her ancilla Eve can obtain half of the data
bits without being detected by the communication parties.

For convenience, we use the same notations as in Ref. \cite {BK}.
We will begin with the property of states
$|\overline{0}\rangle_{12}$ and $|\overline{1}\rangle_{12}$ under
the Pauli operation $\sigma_x=\left(
\begin{array}{c}
0 \quad 1\\
1 \quad 0
\end{array}\right)$
(i.e., flipping). That is, it has the same effect to perform
$\sigma_x$ on the qubit 1 or 2 in the Bell states
$|\overline{0}\rangle_{12}$ and $|\overline{1}\rangle_{12}$, i.e.,
the state $|\overline{0}\rangle_{12}$
($|\overline{1}\rangle_{12}$) will be converted into
$|\overline{1}\rangle_{12}$ ($|\overline{0}\rangle_{12}$) when a
$\sigma_x$ operation is performed on either of the two qubits.
This can be described by
\begin{eqnarray}
\sigma_{x1}\otimes\textbf{I}_2|\overline{0}\rangle_{12}=\textbf{I}_1\otimes\sigma_{x2}|\overline{0}\rangle_{12}=|\overline{1}\rangle_{12}\nonumber\\
\sigma_{x1}\otimes\textbf{I}_2|\overline{1}\rangle_{12}=\textbf{I}_1\otimes\sigma_{x2}|\overline{1}\rangle_{12}=|\overline{0}\rangle_{12}
\end{eqnarray}
where $\textbf{I}$ denotes the identity operation. As a result,
the states $|\overline{0}\rangle_{12}$ and
$|\overline{1}\rangle_{12}$ will be changed into each other under
odd times $\sigma_x$ operation, and be left alone under even times
$\sigma_x$ operation (including the operations on either qubit).
This property will be used later.

Now let us give an explicit description of Eve's strategy. In the
beginning, Alice, Bob and Charlie share a GHZ state
$|G\rangle=(1/\sqrt{2})(|000\rangle+|111\rangle)$ as carrier, and
the data bits Alice wants to distributed to Bob and Charlie can be
represented by $q_1, q_2, q_3, ..., q_n$. Eve prepares a qubit in
state $|0\rangle$ as her ancilla.

(i) In the first round, Eve intercepts the first sending qubit
(the one denoted by subscript 1) and performs a CNOT operation
$C_{1e}$ on this qubit and her ancilla after Alice sent the
particles, which produces the state
\begin{eqnarray}
|\Psi_{abce12}^{0}\rangle=\frac{1}{\sqrt{2}}(|000000\rangle+|111111\rangle) \quad (q_1=0)\nonumber\\
or \quad
|\Psi_{abce12}^{1}\rangle=\frac{1}{\sqrt{2}}(|000111\rangle+|111000\rangle)
\quad (q_1=1)
\end{eqnarray}
and then resends it to Bob. Here we use superscripts 0 and 1 to
denote the states corresponding to $q_1=0$ and $q_1=1$,
respectively. This notation also applies to the following
equations and we will, for simplicity, suppress the word ``or''
later.

It is obvious that these eavesdropping actions introduce no error
when Bob and Charlie disentangle the sending qubits from the
carrier. After that, the state of Alice, Bob, Charlie and Eve can
be specified by
\begin{eqnarray}
|\Theta_{abce}^{0}\rangle_{odd}=\frac{1}{\sqrt{2}}(|0000\rangle+|1111\rangle) \nonumber\\
|\Theta_{abce}^{1}\rangle_{odd}=\frac{1}{\sqrt{2}}(|0001\rangle+|1110\rangle)
\label{eq:second}
\end{eqnarray}

(ii) Eve performs a Hadamard gate on her ancilla. According to the
protocol in Ref. \cite {BK}, before entangle
$|\overline{q}_2\rangle$ to the carrier, Alice, Bob and Charlie
will do the same actions on their respective qubits. As a result,
the entangled state will be converted into
\begin{eqnarray}
|\Theta_{abce}^{0}\rangle_{even}=H^{\otimes4}|\Theta_{abce}^{0}\rangle_{odd}=\frac{1}{2\sqrt{2}}(|0000\rangle+|0011\rangle\nonumber\\+|0101\rangle+|0110\rangle+|1001\rangle+|1010\rangle+|1100\rangle+|1111\rangle) \nonumber\\
|\Theta_{abce}^{1}\rangle_{even}=H^{\otimes4}|\Theta_{abce}^{1}\rangle_{odd}=\frac{1}{2\sqrt{2}}(|0000\rangle-|0011\rangle\nonumber\\-|0101\rangle+|0110\rangle-|1001\rangle+|1010\rangle+|1100\rangle-|1111\rangle)
\label{eq:first}
\end{eqnarray}

(iii) In the second round, Eve intercepts the first sending qubit
and performs a CNOT operation $C_{e1}$ on this qubit and her
ancilla after Alice sent the particles, and then resends it to
Bob. By these actions, Eve's aim is to avoid being detected by the
communication parties.

After Bob and Charlie disentangled the sending qubits from the
carrier, according to the protocol in Ref. \cite{BK}, each of the
four parties has made a CNOT operation on their respective qubit
(as the control qubit) and one of the sending qubits (as the
target qubit). Furthermore, it can be seen from
Eq.~(\ref{eq:first}) that all the items in both
$|\Theta_{abce}^{0}\rangle_{even}$ and
$|\Theta_{abce}^{1}\rangle_{even}$ have even weights. Therefore,
the effect of the above four CNOT operations is equivalent to
flipping either qubit in the transmitted state
$|\overline{q}\rangle$ even times. With the help of the property
of $|\overline{q}\rangle$ introduced in above paragraphs, we can
draw a conclusion that Eve's actions introduce no error in this
round and the state $|\Theta_{abce}^{0}\rangle_{even}$ or
$|\Theta_{abce}^{1}\rangle_{even}$ is not changed. Hence Eve will
avoid the detection of the communication parties, though she can
not obtain any information about the data bit transmitted in this
round.

(iv) As Alice, Bob and Charlie will do, Eve also performs a
Hadamard gate on her ancilla. These four Hadamard operations
change the entangled state $|\Theta_{abce}^{0}\rangle_{even}$
($|\Theta_{abce}^{1}\rangle_{even}$) into
$|\Theta_{abce}^{0}\rangle_{odd}$
($|\Theta_{abce}^{1}\rangle_{odd}$) as described in
Eq.~(\ref{eq:second}), which will be used as a carrier in the next
round.

(v) In the third round, Eve can obtain partial information about
the data bit without being detected by performing operation
$C_{e1}$, making a measurement on the first sending qubit and
performing $C_{e1}$ again.

The particular process is as follows. Suppose the classical bit to
be transmitted (that is, $q_3$) is encoded as $|q,q\rangle$ ($q=0$
or $1$). When Alice has entangled the sending qubits to the
carrier, the state of whole system can be described by
\begin{eqnarray}
|\Phi_{abce12}^{0}\rangle=\frac{1}{\sqrt{2}}(|0000,q,q\rangle+|1111,q+1,q+1\rangle) \nonumber\\
|\Phi_{abce12}^{1}\rangle=\frac{1}{\sqrt{2}}(|0001,q,q\rangle+|1110,q+1,q+1\rangle)
\label{eq:forth}
\end{eqnarray}
where the addition is performed modulo 2.

After the two particles were sent out by Alice, Eve intercepts the
first qubit and performs a CNOT operation $C_{e1}$, producing the
state
\begin{eqnarray}
|\Omega_{abce21}^{0}\rangle=\frac{1}{\sqrt{2}}(|0000,q\rangle+|1111,q+1\rangle)_{abce2}|q\rangle_1 \nonumber\\
|\Omega_{abce21}^{1}\rangle=\frac{1}{\sqrt{2}}(|0001,q\rangle+|1110,q+1\rangle)_{abce2}|q+1\rangle_1
\label{eq:third}
\end{eqnarray}
where we have suppressed the $\otimes$ symbol. It can be seen from
Eq.~(\ref{eq:third}) that the qubit 1 has been disentangled from
the entangled state. Eve then makes a measurement on this qubit in
$B_z=\{|0\rangle, |1\rangle\}$ basis and gets the result $q$ (when
$q_1=0$) or $q+1$ (when $q_1=1$). Let $r_i$ ($i=1,2,...,n$) denote
the eavesdropping results on data bits $q_i$. Eve knows that
$r_3=q_3$ (when $q_1=0$) or $r_3=q_3+1$ (when $q_1=1$).
Afterwards, Eve performs $C_{e1}$ again to recover the state as in
Eq.~(\ref{eq:forth}), and then resends this particle to Bob.

It can be easily verified that Eve's actions introduce no error
when Bob and Charlie disentangle the sending particles from the
carrier. Besides, the carrier and Eve's ancilla is still in the
state $|\Theta_{abce}^{0}\rangle_{odd}$ or
$|\Theta_{abce}^{1}\rangle_{odd}$ as in Eq.~(\ref{eq:second}).

(vi) Eve uses similar operations to eavesdrop in the following
rounds. That is, she takes the same actions as in step.(v) in odd
rounds and the actions as in step.(iii) in even rounds.
Furthermore, when Alice, Bob and Charlie perform Hadamard gates on
their respective qubits at the end of every round, Eve does the
same thing (as in step.(ii) and step.(iv)).

Through above strategy, Eve can avoid the detection of the three
legal parties and even extract information about the odd numbered
data bits at the end of communication. We mean the measurement
results $r_3$, $r_5$, $r_7$, ..., $r_{2m+1}$ (where $m=1,2...$ and
$2m+1\leq n$). Thus Eve can draw a conclusion that Alice's odd
numbered data bits $q_1$, $q_3$, $q_5$, ..., $q_{2m+1}$ must be
equal to $0$, $r_3$, $r_5$, ..., $r_{2m+1}$ or $1$, $r_3+1$,
$r_5+1$, ..., $r_{2m+1}+1$.

The above eavesdropping result seems not so beautiful because
there are still two possibilities. However, according to the
protocol in Ref. \cite{BK}, the three legal parties have to
compare a subsequence of the data bits publicly to detect
eavesdropping, which will leak useful information to Eve. More
specifically, as long as any odd numbered data bit is announced,
Eve can determine which of the two possible results is true. By
this means Eve can obtain the odd numbered data bits completely
except for the little-probability event that all the compared bits
are even numbered.

It should be pointed out that the QKD protocol in Ref. \cite{ZLG},
which inspires the work of Bagherinezhad and Karimipour \cite{BK},
has the same hidden troubles, where Eve can use similar strategy
to extract partial information about the distributed key. From
this kind of attack we can draw two instructive conclusions.
Firstly, when we employ the maximally entangled states as carriers
like the protocols in \cite{BK,ZLG}, it can not effectively
prevent a potential eavesdropping to perform a Hadamard gate
\cite{BK} or a $\pi/4$ rotation \cite{ZLG} on each qubit in the
carriers. Secondly, scheme designers should ensure that the
announcement of subsequence, i.e., the means by which the users
detect eavesdropping, would leak no useful information to the
eavesdropper.

In conclusion, we have presented a eavesdropping strategy which
allows Eve to obtain half of the data bits without being detected
by the communication parties in Bagherinezhad-Karimipour protocol.
Consequently this protocol is insecure against this type of
attack.

This work is supported by the National Natural Science Foundation
of China, Grants No. 60373059; the National Laboratory for Modern
Communications Science Foundation of China, Grants No.
51436020103DZ4001; the National Research Foundation for the
Doctoral Program of Higher Education of China, Grants No.
20040013007; and the ISN Open Foundation.

\end{document}